# Network of vertically c-oriented prism shaped InN nanowalls grown on c-GaN/sapphire template by chemical vapor deposition technique

B.K. Barick*, Rajendra Kumar Saroj, Nivedita Prasad and S. Dhar

*Department of Physics, Indian Institute of Technology, Bombay, Mumbai-400076, India*

**Abstract:**

Networks of vertically c-oriented prism shaped InN nanowalls, are grown on c-GaN/sapphire templates using a CVD technique, where pure indium and ammonia are used as metal and nitrogen precursors. A systematic study of the growth, structural and electronic properties of these samples shows a preferential growth of the islands along [$11\bar{2}0$] and [0001] directions leading to the formation of such a network structure, where the vertically [0001] oriented tapered walls are laterally align along one of the three [$11\bar{2}0$] directions. Inclined facets of these walls are identified as r-planes [($1\bar{1}02$)-planes] of wurtzite InN. Onset of absorption for these samples is observed to be higher than the band gap of InN suggesting a high background carrier concentration in this material. Study of the valence band edge through XPS indicates the formation of positive depletion regions below the r-plane side facets of the walls. This is in contrast with the observation for c-plane InN epilayers, where electron accumulation is often reported below the top surface.

*Corresponding author's E-mail addresses: bkbarick@gmail.com

## I. Introduction

III-V semiconductor nitrides, such as GaN, InN, and AlN, are extensively studied for their applications in electronics and optoelectronics. Among these nitrides, Indium nitride (InN) has the smallest bandgap (~0.7 eV), which makes it potentially interesting for solar cells and infrared detectors [1,2,6-8]. Moreover, theoretical mobility of electrons at room temperature can be as high as 14000 cm$^2$/V·s in InN [3], which is order of magnitude larger than that of GaN (1000 cm$^2$/V·s) [4]. Another important parameter, which is the maximum drift velocity of the electrons at room temperature, can be as high as $4.23 \times 10^7$ cm/s in InN [5] a value that is even larger than that of GaAs and GaN [5]. These properties make InN a potential candidate for applications in high frequency electronics.

However, the growth of epitaxial InN layers is challenging. This is due to several reasons: Firstly, the dissociation temperature of InN is very low, which allows only a narrow temperature window for the growth of this material to take place [9]. Second challenge is the unavailability of suitable substrate materials with sufficiently low lattice and thermal expansion coefficient mismatch with InN. For example, InN has more than 25% lattice mismatch with c-sapphire, which is the commonly used substrates for the growth of III-V nitrides. Even though, it is possible to grow InN epitaxial layers directly on c-sapphire substrates [10-12], the density of dislocations has been found to be much higher than that is typically found in GaN epitaxial films grown on the same substrates [13]. There are efforts to grow InN on AlN, ZnO and GaN templates deposited on sapphire or on Si substrates as the lattice mismatch of InN with these materials is less (~ 10%)[14]. Several groups [15-17] have indeed reported high quality continuous epitaxial InN films on c-GaN templates using molecular beam epitaxy (MBE) technique. Ng et al. have obtained two dimensional (2D) growth of c-InN layers on c-GaN templates by MBE technique at low growth temperatures and low Indium to nitrogen flux ratios.[16] However, high growth temperatures and high indium to nitrogen flux ratios result in the formation of 3D islands. Yamaguchi et al.[18] have reported the epitaxial growth of InN layers on GaN templates by metalorganic vapor phase epitaxy (MOVPE) route. These layers show lower density of structural defects as compared to InN films grown on sapphire and AlN substrates by the same technique. However, it should be noted that lattice mismatch of even 10%, which is typically the mismatch with these template materials, along with a significant thermal expansion coefficient mismatch could be sufficient to introduce a large quantity of strain in the grown InN lattice. Misfit dislocations developed due to strain relaxation can influence the surface diffusion of adatoms and hence can affect the surface morphology in different ways for nitrogen and indium rich conditions. For example, there are reports of epitaxial growth of wedge shaped nanowall network of GaN on c-sapphire [19,20] and Si(111)[21,22] substrates under highly nitrogen rich conditions using MBE. It has been postulated that in nitrogen rich condition, misfit dislocations, which arise as a result of strain relaxation, influence the surface diffusion of adatoms in a way that edges of these defects act as the seed for nucleation of tapered walls [19,20]. This picture can explain the formation of a honeycomb like wall network structure [19,20]. It should be noted that wedge shaped nanowalls made of a polar semiconductor like GaN or InN are particularly interesting as these structures are predicted to have a 2D carrier gas (2DCG) channel formed at the central vertical plane of the wall, which could result in an enhanced carrier mobility[23]. In fact, high mobility and long phase coherence time for the electrons have been recorded in wedged shaped GaN nanowall networks [24,25]. It will be interesting to explore whether such tapered nanowalls of InN can be grown on c-GaN templates utilizing the strain relaxation effect.

In this work we have studied the growth of InN on c-GaN/sapphire templates using a CVD technique, where pure indium and ammonia are used as metal and nitrogen precursors, respectively. Network of vertically *c*-axis oriented prism shaped InN nanowalls, are found to grow on these templates. Structural and electronic properties of these samples are systematically explored. The study shows that a preferential growth of the islands along $[11\bar{2}0]$ and [0001] directions leads to the formation of a network of vertically [0001] oriented tapered walls, which are laterally align along one of the three $[11\bar{2}0]$ directions. The inclined facets of these walls are identified as r-planes [$(1\bar{1}02)$-planes] of wurtzite InN.

## II. Experimental details

InN epitaxial layers were grown by chemical vapor deposition (CVD) method on c-GaN/sapphire template in a tubular furnace with a quartz tube reactor. Prior to the growth, substrates were cleaned by sonicating it successively in Trichloroethylene (TCE), Acetone, and Methanol in an ultrasonic bath for 5 min each before dipping it in HF:H$_2$O (1:10) solution for one minute. Subsequently, the substrates were rinsed in methanol and dried. A quartz boat containing a cleaned substrate placed above the high-pure indium metal (99.99%) was kept at the center of the furnace. The quartz tube was evacuated and purged several times with high pure argon (99.999%) at 200 standard cubic centimeters per minute (sccm) for 20 min and later the rate ($\phi_{Ar}$) was adjusted to a certain value and allowed to continue till the end of the growth process. Temperature of the furnace was ramped at a rate of 22 °C/min until it reached the desired growth temperature $T_G$. When the furnace temperature reached 200 °C, ammonia (99.999%) was allowed to flow into the reactor at a certain flow rate $\phi_{NH_3}$ and continued till the end of the growth. The flow rates of argon and ammonia into the reactor were controlled through mass flow controllers (MFC). $T_G$ was maintained at a fixed value for certain growth time $t_G$ before the cooling process was initiated. The furnace was allowed to cool naturally to the room temperature. When the furnace was cooled down to 200 °C, ammonia was switched off. Sample was taken out from the reactor at room temperature under the flow of argon. Several samples were grown with growth time $t_G$ ranging from 1 to 5 h at $T_G$ of 550 °C, with argon and ammonia flow rates of 100 and 40 sccm, respectively.

High resolution X-ray diffraction experiments were performed using a Rigaku SmartLab diffractometer utilizing Cu K$_\alpha$ radiation. Wide angle $\omega - 2\theta$ scans were carried out for $2\theta$ ranging from 20° to 80°. To understand the orientation and epitaxial quality, in-plane ($\phi$-scan) and out-of-plane ($\omega$-scan) measurements were carried out for $(10\bar{1}0)$ and $(0002)$ reflections, respectively. Here, the $\phi$-scan of $(10\bar{1}0)$ plane was performed in in-plane geometry by keeping $2\theta_\chi = 29.4°$ [Bragg angle for $(10\bar{1}0)$ reflection] and lifting source and a wide open detector up by a small angle of 0.25°. The surface morphology for all these samples was analyzed by scanning electron microscopy (SEM). Band-edges of the InN samples were estimated from optical-absorption measurements using Lambda 950 UV/Vis/NIR spectrophotometer. Raman spectra were recorded over a scan range of 100–800 cm$^{-1}$ in a backscattering geometry using 514.5 nm line of an Ar-ion laser. XPS study of these samples was carried out using MULTILAB from Thermo VG Scientific using Al K-α line as the x-ray source. Binding energies were calibrated using carbon C1s peak (284.6eV) as a reference standard.

### III. Results and discussion:

#### 1. Surface Morphology

Figure 1 compares the top view SEM images of the samples grown for (a) 1 h, (b) 3 h and (c) 5 h. In case of one hour grown sample, formation of faceted islands are quite evident. Many of these islands are hexagonal pyramid shaped, suggesting that they are grown along [0001] direction. In fact, symmetric x-ray diffraction (XRD) study (discussed later in Fig. 2) confirms the [0001] vertical orientation of the islands. Furthermore, in-plane XRD $\phi$-scans for ($11\bar{2}0$) and ($1\bar{1}00$) planes are carried out on these samples in order to investigate the in-plane crystalline orientation, which has been discussed later in Fig.3. This study reveals that the base lines of the facets for all these pyramidal mounds are oriented perpendicular to [$1\bar{1}00$] directions of wurtzite InN lattice. A careful inspection also reveals that the islands are elongated along one of the three [$11\bar{2}0$] directions. This is quite evident from the inset of the panel (a) where an elongated island is shown in an extended scale. In case of 3h [panel (b)] and 5h [panel (c)] grown samples, only connected network of tapered walls can only be seen. Cross sectional view of these walls is shown in the inset of panel (b). Triangular cross-sections of the walls is evident from the figure. Both the side facets of a wall are measured to be inclined at an angle of ~58° with respect to the substrate surface. In fact, this inclination angle is found to be the same for all the walls investigated here. Note that the angle between r- and c-plane of wurtzite InN is 58°. This finding suggests that the side facets of the tapered walls must be r-plane ($1\bar{1}02$) of wurtzite InN. From panel (b) and (c) it is clear that these tapered walls are elongated along one of the three [$11\bar{2}0$] directions, which is more evident from the inset (ii) of panel (c) of the figure representing the fast Fourier transform (FFT) of the image of panel (c), which shows three clear maximum-intensity-lines oriented perpendicular to the three [$11\bar{2}0$] crystalline directions marked with arrows in the image. This finding along with the inset of panel (a) suggest that the islands are elongated along [$11\bar{2}0$] directions. This indicates a faster growth along one of the three [$11\bar{2}0$] directions as compared to other lateral directions. From these figures, it appears that at the beginning, these nucleation cites elongate themselves along one of the three [$11\bar{2}0$] directions, which is randomly decided [see panel (a)]. Interestingly, however, on the substrate plane these islands continue to grow along the direction of their initial elongation resulting in the formation of long islands [Inset of panel (b)]. This type of lateral growth along with the vertical growth lead to the formation of a network of wedge shaped walls. It should be noted that the side facets of these islands are found to be r-planes. It is plausible that the growth on these planes is much slower than [$11\bar{2}0$] and c-directions. This might be the reason for the growth of prism shaped islands.

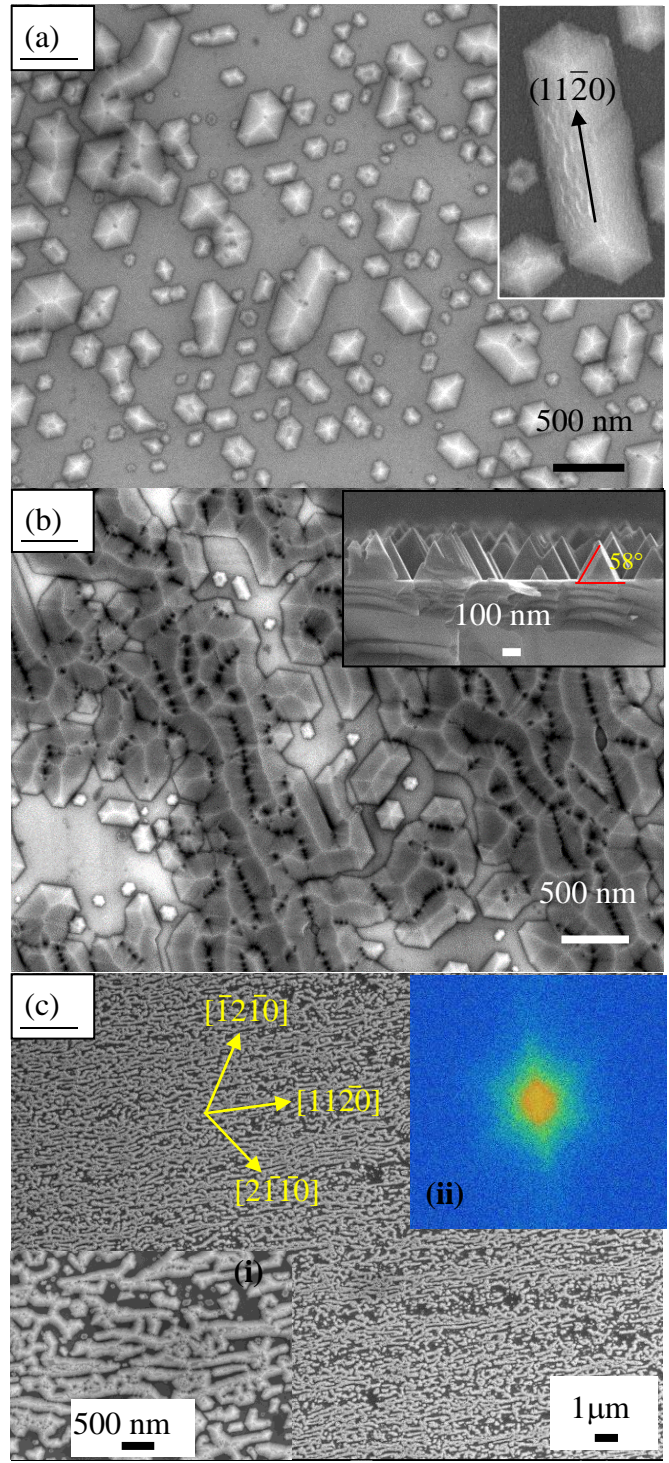

**Figure 1** Top view SEM images of the samples grown for (a) 1 h (b) 3 h (c) 5 h, Insets of panel (a) shows the expanded SEM image of an elongated island, inset of panel (b) shows the cross sectional SEM image of the sample, inset (i) of panel (c) shows the top view SEM images recorded at higher magnification of the samples and inset (ii) of panel (c) shows fast Fourier transform (FFT) of the image of panel (c).

## 2. Structural Studies

Figure 2 shows the high-resolution wide-angle XRD (ω-2θ) profiles for InN grown on GaN/sapphire templates for different growth times. In all cases, peaks at 31.3° and 65.4°, which correspond to (0002) and (0004) reflections for the wurtzite phase of InN are present. Reflections related to any other planes could not be detected. This clearly suggests an epitaxial growth of InN layers along c-direction.

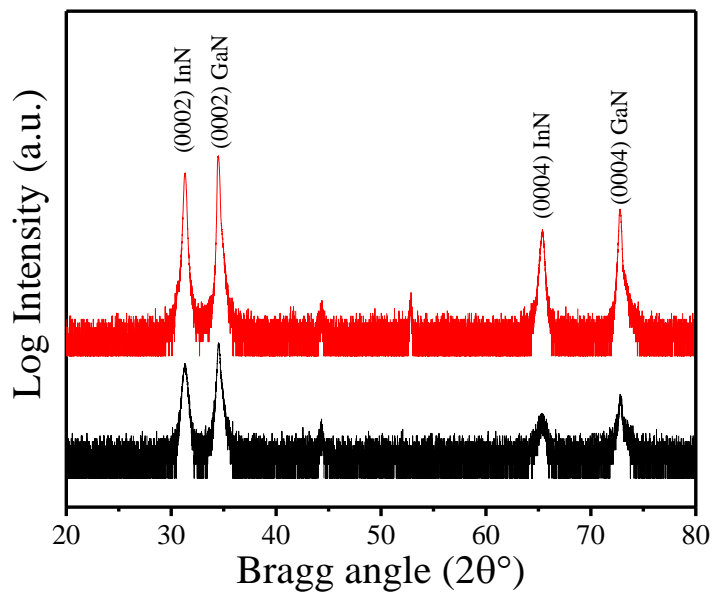

**Figure 2** High-resolution wide angle XRD ($\omega - 2\theta$) profiles for InN grown on c-GaN/sapphire templates for 1h (black solid line) and 5h (blue solid line).

Figure 3 shows the rocking curve for (0002) reflection recorded for the 5 h grown sample. Full width at the half maximum (FWHM) for the rocking curves is found to be 0.27°, which is comparable to that is reported for our epitaxial InN layers grown on c-sapphire substrates [9]. Average angle of tilt $\alpha_{ti}$ as well as the lateral coherence length $L_I$ for the grains, which are in this case of nanowalls, can be estimated using the relation [26,27]:

$$\beta \sin\theta / \lambda = \alpha_{ti} \sin\theta / \lambda + 0.9 / 2L_I \qquad (1)$$

where $\beta$ and $\theta$ are the FWHM and the Bragg angles associated with the rocking curves for (0002), (0004) and (0006) reflections. $\beta \sin\theta / \lambda$ Versus $\sin\theta / \lambda$ plot, which is also called the Williamson Hall plot, is shown in the inset of the figure. A straight-line fit yields $\alpha_{ti} = $ 0.0046 rad and $L_l$ = 954 nm. It has to be noted that, $\alpha_{ti}$ and $L_l$ are estimated to be 0.0103 rad and 302 nm, respectively, for InN epitaxial layers grown on sapphire substrates at similar growth conditions [10].

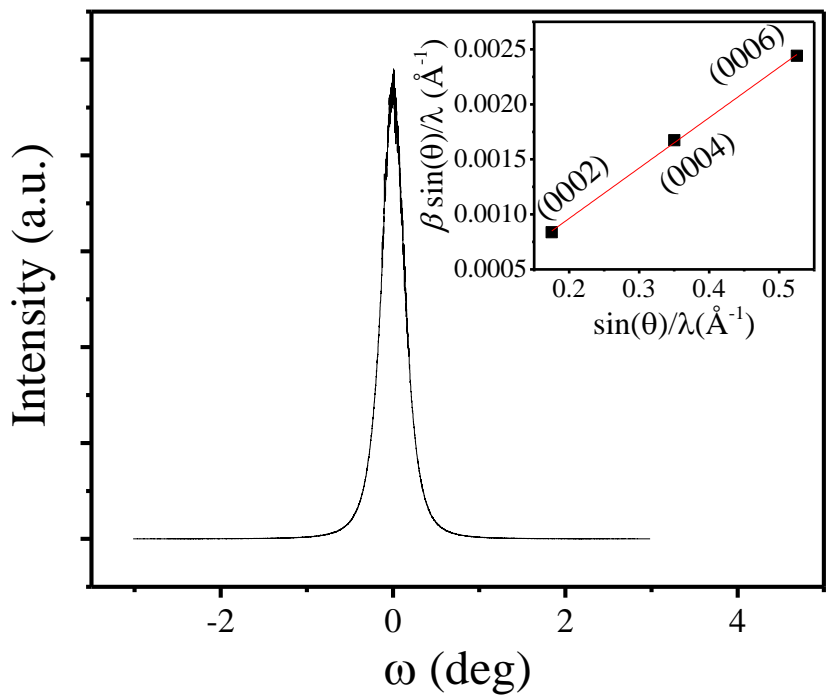

**Figure 3** Rocking curve for the (0002) reflection. Inset shows the Williamson Hall plot for the same and its higher order reflections.

Figure 4(a) shows the wide angle $\phi$-scan for ($10\bar{1}0$) plane performed in the in-plane geometry for the same sample. Appearance of six equidistant peaks suggests a good in-plane crystalline orientation of the walls. Unequal intensity of $\phi$-scan features may indicate a non-uniform distribution of the nanowall density over the surface. Figure 4(b) shows the ($10\bar{1}0$) $\phi$- rocking curve for the same sample. FWHM of the peak is found to be 0.67°. Average angle of twist $\alpha_{tw}$ and coherence length along c-direction $L_v$ for the nanowalls can be obtained from the Williamson Hall plot for ($10\bar{1}0$), ($20\bar{2}0$) and ($30\bar{3}0$) reflections following a similar equation as eq. (1). Such a plot is shown in the inset of Fig. 4(b). Fitting the data with a straight-line yields $\alpha_{tw}$ = 0.01 rad. and $L_v$ =250 nm. It has to be mentioned that, $\alpha_{tw}$ and $L_v$ are estimated to be 0.026 rad and 677 nm, respectively, for InN epitaxial layers grown on sapphire substrates at similar growth conditions [10].

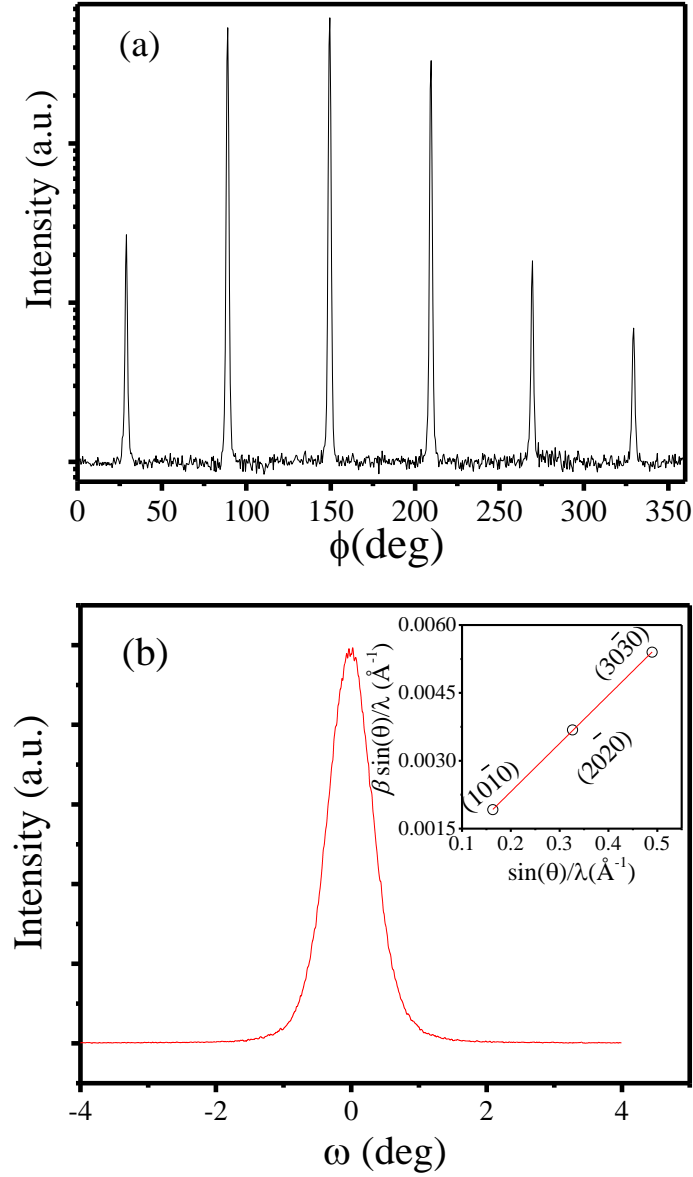

**Figure 4** (a) Wide angle $\phi$-scan for $(10\bar{1}0)$ planes performed in the in-plane geometry for the same sample. (b) $\phi$-rocking curve ($\phi$-scan within a narrow range). Inset of (b) shows the Williamson Hall plot for $(10\bar{1}0)$ and its higher order reflections.

Strain in the grown structure can be obtain from the FWHM associated with the $\omega-2\theta$ and $2\theta_\chi-\phi$ scans recorded for the symmetric and the vertical planes respectively following a relation, which is similar to Eqn. (1) [28].

$$\beta\cos\theta/\lambda = 4\varepsilon\sin\theta/\lambda + 0.9/D \qquad (2)$$

where $\beta$ and $\theta$ are the FWHM and the Bragg angles associated with the $\omega-2\theta$ ($2\theta_\chi-\phi$) scans for the symmetric(vertical) plane and its higher order reflections. $\varepsilon_l$ ($\varepsilon_v$) is the lateral
9

(vertical) strain in the lattice. *D* is the mean crystallite size. Figure 5 shows such Williamson Hall plots for (0002) plane (symmetric plane) and its higher order reflections (Plot-I) as well as for ($10\bar{1}0$) plane (vertical plane) and its higher order reflections (Plot-II). A straight-line fit of Plot-I returns the strain along the c-direction to be $\varepsilon_v$ = 0.00056 and crystallite size of 41 nm. A similar analysis of Plot-II yields the lateral strain to be $\varepsilon_l$ = 0.000355 and crystallite size of 20 nm. Note that, $\varepsilon_v$ and $\varepsilon_l$ are estimated to be 0.00103 and 0.00164 for the InN epitaxial layer grown on sapphire substrate at similar growth conditions [10].

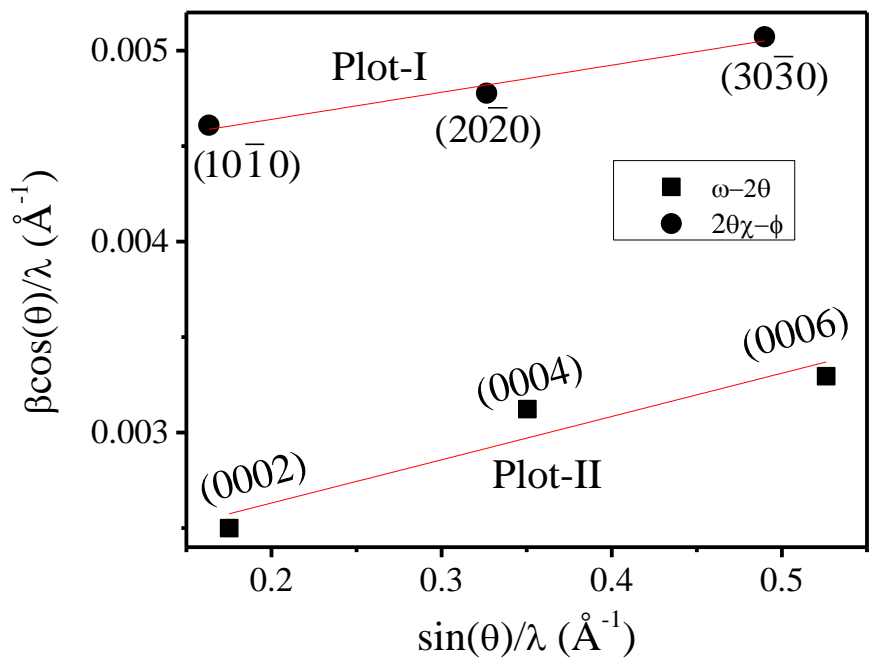

**Figure 5** Williamson Hall plots for (0002) and its higher order reflections obtained from the $\omega-2\theta$ scans (Plot-I) as well as for ($10\bar{1}0$) and its higher order reflections obtained from the $2\theta_\chi-\phi$ scans (Plot-II).

## 3. Electronic properties

Figure 6 (a) compares the Raman spectra recorded for a bare c-GaN/sapphire template and the InN nanowalls grown on the same template. Raman spectrum for an InN epitaxial layer grown on c-sapphire substrate by the same technique is also shown in the figure. $E_2$ (high) and $A_1$(LO) characteristic modes for InN are clearly observed at 489.6 cm$^{-1}$[29] and 590 cm$^{-1}$[30], respectively. Absence of any other Raman feature in the back scattered Raman

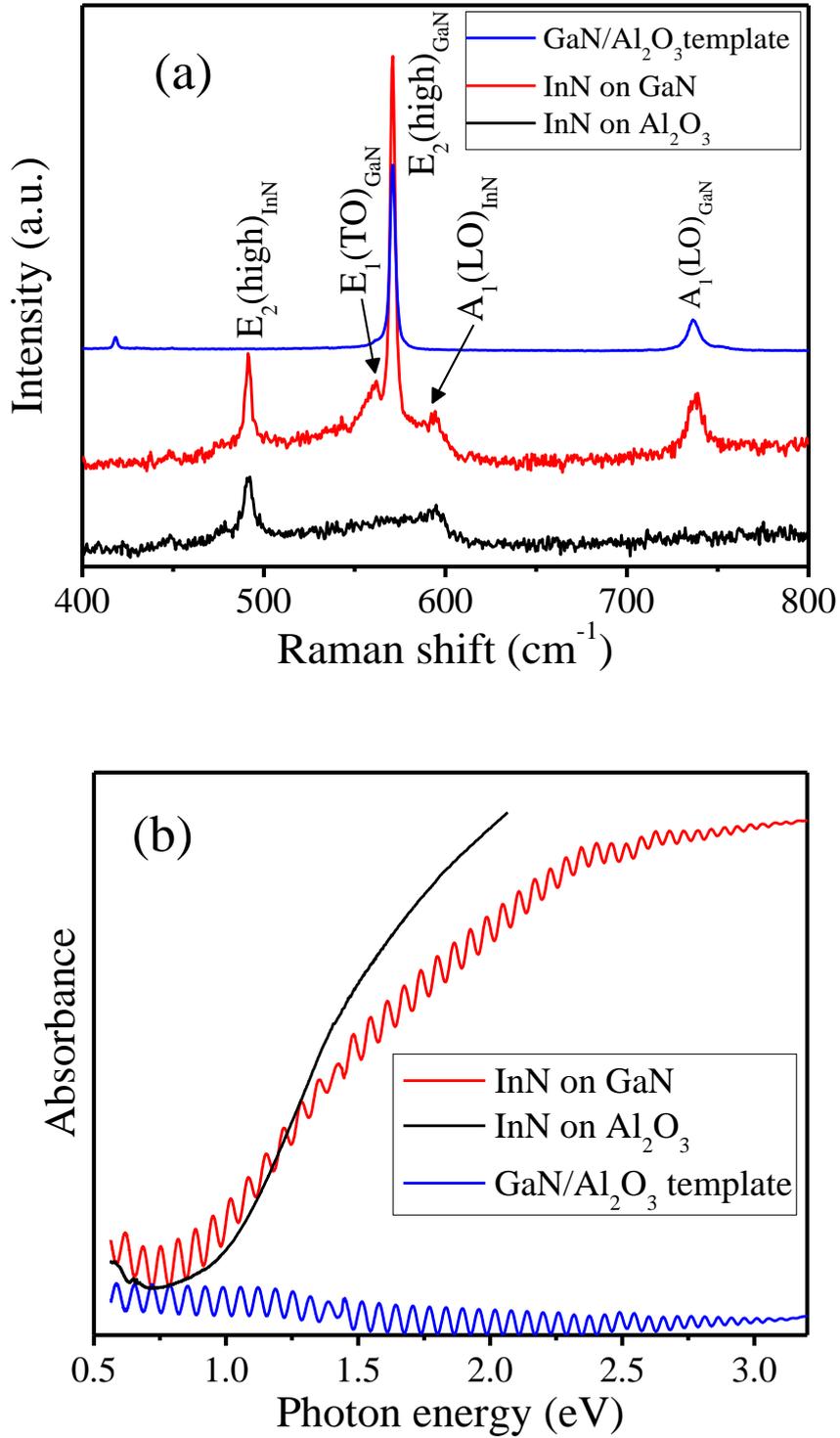

**Figure 6** (a) Raman spectra for the InN nanowall sample grown on c-GaN/sapphire template for 5h (red solid line), a bare GaN/Al$_2$O$_3$ template (blue solid line) and an InN epitaxial layer grown on sapphire substrate by the same technique (black solid line) (b) Photo absorption profiles for these samples.

spectra for both the InN samples strongly suggest a c-directional growth of the material in both the cases, which is consistent with the observations of Fig. 2 and 4. Figure 6 (b) compares the photo-absorption profiles for the same three samples. It clearly shows an onset of absorption at around 1 eV for both InN samples. The oscillation observed in the absorption profile of InN grown on c-GaN/sapphire template corresponds to the interference of light at the surface and interface of the GaN layer. Note that the same oscillation is observed in case of the bare template as well. Blue shift of the absorption edge with respect to the band gap of 0.7 eV [2] of InN suggest a degenerate nature both for the InN nanowalls and the continuous layer.

Figure 7 shows XPS spectra recorded at the valence band edge for the InN nanowall. Note that the zero energy corresponds to the position of the surface Fermi energy. At the surface, the valence band maximum is estimated to be located at 0.4 eV below the surface Fermi level ($E_{FS}$). InN has a band gap of 0.7 eV at room temperature [29], which means that the surface Fermi level is laying 0.3 eV below the conduction band minimum ($E_C$) in these samples as shown schematically in the inset of Fig 7. Position of Fermi level below the conduction band suggests depletion of electron on the surface of nanowall. It should be noted that, in case of c-InN epitaxial layers grown on c-sapphire substrates, the conduction band minimum has been found to lay below the fermi level at the surface. In fact, similar result has been reported by many groups for c-InN epitaxial films [31,32]. Note that the side facets of these walls are identified as r-planes (see Fig.1). While the c-surface of InN is associated with a surface accumulation of electrons, r-surface results in the formation of a depletion region below the surface. Note that the effect of surface accumulation of electrons in case of c-surface has been explained in terms of certain donor like surface states, which might be resulting either from the reconstruction of the dangling bonds or adsorption of impurity ions/molecules on the c-surface [33,34]. One plausible reason for the formation of depletion regions below the r-plane side facets of the walls is the spontaneous polarization effect, which can attribute a net polarization of $P\cos\theta_r$ on the side facets, where $P$ is the polarization and $\theta_r = 58^o$, the angle between the r- and c-planes. $P$ = -0.032 C/m$^2$ for indium polar InN [35]. Considering, In-polar growth of the walls, one can expect polarization induced negative charges on the side facets of the walls. These negative charges can push the electron cloud of the unintentionally n-type InN nanowalls towards the center resulting in the formation of positive depletion regions below the surface of the walls. This scenario has been theoretically investigated in case of c-axis oriented wedge shaped GaN nanowalls [23]

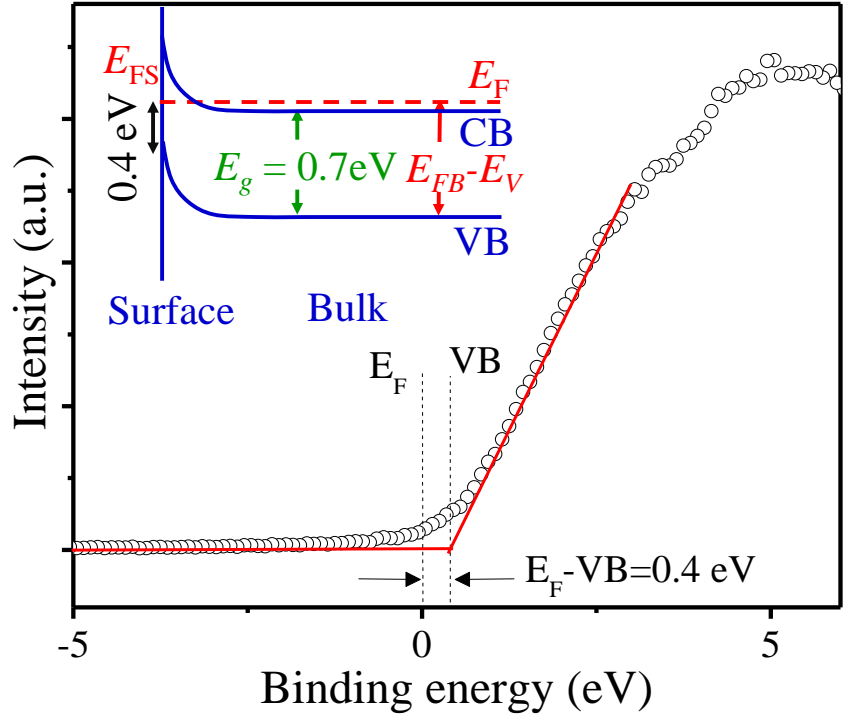

**Figure 7** XPS spectrum recorded at the valence band edge for InN nanowalls grown on GaN/Al$_2$O$_3$ substrate for 5h. Inset: schematic representation of the band bending at the surface.

**IV. Conclusion:**

Network of vertically *c*-axis oriented prism shaped InN nanowalls, which are elongated along one of the six [11$\bar{2}$0] directions, can be grown on c-GaN/sapphire templates. The side facets of these walls are found to be r-planes of wurtzite InN. Bandgap of the film is observed to be Burstein-Moss shifted, which is likely to be due to high background electron concentration. XPS studies suggest the formation of positive depletion regions below the r-plane side facets of the walls. Note that the finding is different from c-plane InN epilayers, where electron accumulation is often reported below the top surface.

**Acknowledgements**

We acknowledge Sophisticated Analytical Instrument Facility (SAIF), IIT Bombay, Center of excellence for nanoelectronics (CEN) for providing various experimental facilities and FIST (Physics)-IRCC central SPM facility of IIT Bombay. This work was supported by Department of Science and Technology (DST) under Grant No: SR/S2/CMP–71/2012 and Council of Scientific & Industrial Research (CSIR) under Grant No: 03(1293)/13/EMR-II, Government of India.

**References**

1. E. Trybus, G. Namkoong, W. Henderson, S. Burnham and W. A. Doolittle, InN: A material with photovoltaic promise and challenges, J. Cryst. Growth, 288, (2006) 218.

2. B. Tekcan, S. Alkis, M. Alevli, N. Dietz, B. Ortac, N. Biyikli and A. K. Okyay, A Near-Infrared Range Photodetector Based on Indium Nitride Nanocrystals Obtained Through Laser Ablation, IEEE Electron Device Lett., 35 (2014) 936.

3. V. M. Polyakov and F. Schwierz, Low-field electron mobility in wurtzite InN, Appl. Phys. Lett., 88 (2006) 032101.

4. U. V. Bhapkar and M. S. Shur, Monte Carlo calculation of velocity-field characteristics of wurtzite GaN, J. Appl. Phys., 82 (1997) 1649.

5. S. K. O'Leary, B, E. Foutz, M. S. Shur and L. F. Eastman, Steady-state and transient electron transport within bulk wurtzite indium nitride: An updated semiclassical three-valley Monte Carlo simulation analysis, Appl. Phys. Lett., 87 (2005) 222103.

6. D. V. P. McLaughlin and J. M. Pearce, Progress in indium gallium nitride materials for solar photovoltaic energy conversion, Metall. Mat. Trans. A, 44 (2013) 1947.

7. C. Rivera, J. Pereiro, Á. Navarro, E. Muñoz, O. Brandt and H. T. Grahn, Advances in Group-III-Nitride Photodetectors, TOEEJ 4 (2010) 1.

8. Z. H. Zhang, W. Liu, Z. Ju, S. T. Tan, Y. Ji, Z. Kyaw, X. Zhang, L. Wang, X. W. Sun and H. V. Demir, InGaN/GaN multiple-quantum-well light-emitting diodes with a grading InN composition suppressing the Auger recombination, Appl. Phys. Lett., 105 (2014) 033506.

9. V. Woods and N Dietz, InN growth by high-pressures chemical vapor deposition: Real-time optical growth characterization, Mater. Sci. Eng. B, 127 (2006) 239.

10. B. K. Barick, N. Prasad, R. K. Saroj, S. Dhar, Structural and electronic properties of InN epitaxial layer grown on c-plane sapphire by chemical vapor deposition technique, J. Vac. Sci. Technol. A, 34 (2016) 051503.

11. W. K. Chen, Y. C. Pan, H. C. Lin, J. Ou, W. H. Chen and M. C. Lee, Growth and X-ray Characterization of an InN Film on Sapphire Prepared by Metalorganic Vapor Phase Epitaxy, Jpn. J. Appl. Phys. 36 (1997) L1625.

12. Q. X. Guo, T. Yamamura, A. Yoshida and N. Itoh, Structural properties of InN films grown on sapphire substrates by microwave excited metalorganic vapor-phase epitaxy, J. Appl. Phys., 75 (1994) 4927.

13. W. Cao, J. M. Biser, Y. K. Ee, X. H. Li, N. Tansu, H. M. Chan and R.P. Vinci, Dislocation structure of GaN films grown on planar and nano-patterned sapphire, J. Appl. Phys., 110 (2011) 053505.

14. A. G. Bhuiyan, A. Hashimoto and A. Yamamoto, Indium nitride (InN): A review on growth, characterization, and properties, J. Appl. Phys., 94 (2003) 2779.

15. W. C. Chen, S. Y. Kuo, W. L. Wang, Jr S. Tian, W. T. Lin, F. I. Lai and L. Chang, Study of InN epitaxial films and nanorods grown on GaN template by RF-MOMBE, Nanoscale Res. Lett., 7 (2012) 468.

16. Y. F. Ng, Y. G. Cao, M. H. Xie, X. L. Wang and S. Y. Tong, Growth mode and strain evolution during InN growth on GaN(0001) by molecular-beam epitaxy, Appl. Phys. Lett., 81 (2002) 3960.

17. E. Dimakis, E. Iliopoulos, K. Tsagaraki, Th. Kehagias, Ph. Komninou and A. Georgakilas. Heteroepitaxial growth of In-face InN on GaN (0001) by plasma-assisted molecular-beam epitaxy, J. Appl. Phys., 97 (2005) 113520.

18. S. Yamaguchi, M. Kariya, S. Nitta, T. Takeuchi, C. Wetzel, H. Amano and I. Akasaki, Structural properties of InN on GaN grown by metalorganic vapor-phase epitaxy, J. Appl. Phys., 85 (1999) 7682.

19. M. Kesaria, Satish Shetty and S. M. Shivaprasad, Evidence for dislocation induced spontaneous formation of GaN nanowalls and nanocolumns on bare c-plane sapphire, Cryst. Growth Des., 11 (2011) 4900.

20. V. Thakur, M. Kesaria and S. M. Shivaprasad, Enhanced band edge luminescence from stress and defect free GaN nanowall network morphology, Solid State Commun., 171 (2013) 8.

21. A. Zhong and K. Hane, Growth of GaN nanowall network on Si (111) substrate by molecular beam epitaxy, Nanoscale Res. Lett., 7 (2012) 686.

22. A. Zhong and K. Hane, Characterization of GaN nanowall network and optical property of InGaN/GaN quantum wells by molecular beam epitaxy, Japan. J. Appl. Phys., 52 (2013) 08JE13.

23. S. Deb, H. P. Bhasker, V. Thakur, S. M. Shivaprasad and S. Dhar, Polarization induced two dimensional confinement of carriers in wedge shaped polar semiconductors, Sci. Rep., 6 (2016) 26429.

24. H. P. Bhasker, S. Dhar, A. Sain, M. Kesaria and S. M. Shivaprasad, High electron mobility through the edge states in random networks of c-axis oriented wedge-shaped GaN nanowalls grown by molecular beam epitaxy. Appl. Phys. Lett., 101 (2012) 132109.

25. H. P. Bhasker, V. Thakur, S. M. Shivaprasad and S. Dhar, Quantum coherence of electrons in random networks of c-axis oriented wedge-shaped GaN nanowalls grown by molecular beam epitaxy. J. Phys. D: Appl. Phys., 48 (2015) 255302.

26. M. A. Moram and M. E. Vickers, X-ray diffraction of III-nitrides, Rep. Prog. Phys., 72 (2009) 036502.

27. L. Tarnawska, A. Giussani, P. Zaumseil, M. A. Schubert, R. Paszkiewicz, O. Brandt, P. Storck and T. Schroeder, Single crystalline $Sc_2O_3$/$Y_2O_3$ heterostructures as novel engineered buffer approach for GaN integration on Si (111), J. Appl. Phys., 108 (2010) 063502.

28. G. K. Williamson, W. H. Hall: Acta Metall., 1 (1953) 22.

29. T. Matsuoka, H. Okamoto, M. Nakao, H. Harima and E. Kurimoto, Optical bandgap energy of wurtzite InN, Appl. Phys. Lett., 81 (2002) 1246.

30. N. Domènech-Amador, R. Cuscó, and L. Artús, T. Yamaguchi and Y. Nanishi, Raman scattering study of anharmonic phonon decay in InN, Phys. Rev. B, 83 (2011) 245203.

31. R. P. Bhatta, B. D. Thoms, A. Weerasekera, A. G. U. Perera, M. Alevli, and N. Dietz, Carrier concentration and surface electron accumulation in indium nitride layers grown by high pressure chemical vapor deposition, J. Vac. Sci. Technol. A, 25 (2007) 967

32. K. A. Rickert, A. B. Ellis, F. J. Himpsel, H. Lu, W. Schaff, J. M. Redwing, F. Dwikusuma and T. F. Kuech, X-ray photoemission spectroscopic investigation of surface treatments, metal deposition, and electron accumulation on InN, Appl. Phys. Lett., 82 (2003) 3254.

33. B. K. Barick and S. Dhar, Photoinduced electromotive force on the surface of InN epitaxial layers, arXiv:1706.05359

34. I. Mahboob, T. D. Veal, L. F. J. Piper, and C. F. McConville, Hai Lu and W. J. Schaff, J. Furthmüller and F. Bechstedt, Origin of electron accumulation at wurtzite InN surfaces, Physical Review B 69 (2004) 201307(R).

35. F. Bernardini, V. Fiorentini and D. Vanderbilt, Spontaneous polarization and piezoelectric constants of III-V nitrides, Phys. Rev. B, 56 (1997) R10024.